# Open Data Analytical Model for Human Development Index Optimization to Support Government Policy

Andry Alamsyah, Tieka T. Gustyana, Adit D. Fajaryanto, Dwinanda Septiafani
School of Economics and Business, Telkom University, Bandung, Indonesia
andrya@telkomuniversity.ac.id

**Abstract:** The transparency nature of *Open Data* is beneficial for citizens to evaluate government work performance. In Indonesia, each government bodies or ministry have their own standard operation procedure on data treatment resulting in incoherent information between agent and likely to miss valuable insight. Therefore, our motivation is to show the advantage of *Open Data* movement to support unified government decision making. We use dataset from data.go.id which publish official data from each government bodies. The idea is by using those official but limited data, we can find important pattern. The case study is on *Human Development Index* value prediction and its clustered nature.

We explore the data pattern using two important data analytics methods classification and clustering procedure. Data analytics is the collection of activities to reveal unknown data pattern. Specifically, we use *Artificial Neural Network* classification and *K-means* clustering. The classification objective is to categorize different level of *Human Development Index* of cities or region in Indonesia based on *Gross Domestic Product*, *Number of Population in Poverty*, *Number of Internet User, Number of Labors* and *Number of Population* indicators data. We determined which city belongs to four categories of Human Development stated by UNDP standard. The clustering objective is to find the group characteristics between *Human Development Index* and *Gross Domestic Product*.

## 1. Introduction

*Open data,* especially open government data, is a tremendous resource that is yet fully utilized in many countries. *Open data* have many advantages both for citizens or government itself, for example the citizens could be more aware about what their government work performances or the government services could be significantly improved as operational data becomes available to improve business processes and shorten delivery times. Inconsistency in policy and utilization of data storage in Indonesia detain development synergy by government. According to the agenda of the *Millennium Development Goals* (MDGs) by the *United Nations Development Program* (UNDP) on the goal to 2017 'Strengthen the means of implementation and revitalize the global partnership for sustainable development' points to 2018 stated that in 2020, all developing countries are expected to increase significantly the availability of high-quality, timely and reliable data (United Nations, 2016). In Indonesia case, each government bodies have their own standard operation procedure on data collecting, that makes incompatibility data about a particular case. For example, in the case of Indonesian food reliability strategy, we need high quality, reliable, accurate, and timely manner data that can describe and predict the when the food shortage will happen, what area need to be strengthened of, what strategy to keep the balance of food availability, and other issues (Tempo, 2016). This is corroborated by the statement of the President of the Republic of Indonesia and Coordinating Minister for the Economy (CNN Indonesia, 2016). This condition led policies by government to missteps. The unity of the data can be done if the quality of the data from each government bodies is adequate by applying the regulations to fix the infrastructure data management and public information services of government, yet this still not optimized by government (Kebebasan Informasi, 2016).

*Human Development Index* (HDI) is a to measure and evaluate Indonesia government policy (Biro Pusat Statistik, 2016). According to the UNDP, human development is defined as expansion option for citizen to have choices. It means as the efforts towards "expansion options" as well as the extent achieved from these efforts. At the same time the human development is as the formation of human capabilities through improved level health, knowledge, and skills; as well as utilization ability/skills. The concept of development over much broader sense than the concept of economic development that emphasizes the growth (including economic growth), basic needs, community welfare, or human resource development (Alkire, 2010).

Data Analytics is collection of methods to measure human characteristic or behavior based on available data or the digital trace left in the internet (Liu, 2009). There are many method to finding human behavioral pattern, one of them is data mining. Data mining provide many model to fit the data pattern such as regression, classification, association, outlier detection, time series, clustering, and many other models. However, since the nature of Indonesia *Open Data* is limited, in this case is incomplete data, then we can only explore on limited models, which are classification and clustering.

In this paper, we show how we can predict HDI value and finding pattern of HDI grouping pattern with *Gross Domestic Product* (GDP). We use the best and suitable classification and clustering technique to the Open Data limited format. The techniques are *Artificial Neural Network* (ANN) classification and *K-Means* clustering. ANN is used because the ability to find all the possibilities for relations between all variables or indicators measured. K-means is used because of its simplicity and effectiveness to find clusters in data.

## 2. Theoretical Background and Data Characteristic

Data mining is the exploratory process and analysis from large quantities of data. Data source can come from database, data warehouse, web, other information repository, and data streaming (Han et al, 2012). Two data mining model used in this paper are classification and clustering model. Classification model is grouping the data based on historical behavior of the previous data. The observed variables are labeled to support the automation classification process (Tan et al, 2014). Clustering models refers to data grouping process into classes of similar data. The similarity procedure done by unsupervised fashion, there are methods such as distance to measure the similarity automatically (Larose, 2015).

*Artificial Neural Network* (ANN) is a classification technique that imitate the way of human nervous system working. They use for a wide variety of task, from relatively simple classification problems to speech recognition and computer vision (Kriesel, 2005). *K-means clustering* algorithm is a simple and effective algorithm for finding cluster in data. *K-means* split data into *k* cluster that have been predetermined in advance (Larose, 2015).

We use *Open Data* that provide free datasets about many variables measured in connection with Indonesia development. The datasets are available to download in One Data Indonesia project on the www.data.go.id website. One Data Indonesia is an Indonesian Government project to make centralized and open website that contains data from all ministry departments in Indonesia as a part of government commitment on *Open Government Partnership*. At this moment One Data Indonesia is still a pilot project, Indonesia Government targeting the website are fully ready by 2018. Preview of the data can be seen in Figure 1. Here we can see that the data is incomplete or missing for certain measurement indicators, and for some time stamp measurement (yearly). Because of this obstacle, we use only the complete data for model construction shown in Table 1.

| 971 | Area Name | Indicator Name | 2005 | 2006 | 2007 | 2008 | 2009 | 2010 | 2011 | 2012 | 2013 | 2014 |
|---|---|---|---|---|---|---|---|---|---|---|---|---|
| 972 | Bandung, Kota | Length of Province Road: Fair (in km) (BPS Data, Province only) | | | | | | | | | | |
| 973 | Bandung, Kota | Length of Province Road: Good (in km) (BPS Data, Province only) | | | | | | | | | | |
| 974 | Bandung, Kota | Length of Province Road: Gravel (in km) (BPS Data, Province only) | | | | | | | | | | |
| 975 | Bandung, Kota | Length of Province Road: Light Damage (in km) (BPS Data, Province only) | | | | | | | | | | |
| 976 | Bandung, Kota | Length of Province Road: Other (in km) (BPS Data, Province only) | | | | | | | | | | |
| 977 | Bandung, Kota | Literacy Rate for Population age 15 and over (in % of total population) | 98.91 | 99.58 | 99.57 | 99.63 | 99.54 | 99.65 | 99.09 | 99.72 | 99.56823 | 99.96 |
| 978 | Bandung, Kota | Monthly Per Capita Household Education Expenditure (in IDR) | 31319 | 33872 | 36037 | | 69869 | 65738 | 88087 | 60105 | 96346.31 | 125040.05 |
| 979 | Bandung, Kota | Monthly Per Capita Household Health Expenditure (in IDR) | 10196 | 7750 | 14952 | | 19397 | 18786 | 30317 | 32602 | 53136.06 | 54608.01 |
| 980 | Bandung, Kota | Monthly Per Capita TOTAL Household Expenditure for The Poorest 20 percent (in ID | 167069 | 200046 | 214100 | | 295634 | 319378 | 316424 | 293831 | 383063.6 | 389530.58 |
| 981 | Bandung, Kota | Morbidity Rate (in %) | 22.45014 | 16.29512 | 24.29832 | 28.826 | 32.57811 | 31.88825 | 21.7686 | 24.25203174 | 26.8817693 | |
| 982 | Bandung, Kota | Net Enrollment Ratio: Junior Secondary (in %) | 71.2 | 75.14 | 66.47 | 70.16 | 71.82 | 60.27 | 72.76 | 71.43 | 89.093185 | 89.85 |
| 983 | Bandung, Kota | Net Enrollment Ratio: Primary (in %) | 89.19 | 90.41 | 92.07 | 89.97 | 97.66 | 93.93 | 89.16 | 92.92 | 97.576813 | 97.95 |
| 984 | Bandung, Kota | Net Enrollment Ratio: Senior Secondary (in %) | 51.43 | 58.99 | 58.05 | 59.22 | 45.4 | 54.43 | 52.8 | 62.67 | 51.177273 | 59.76 |
| 985 | Bandung, Kota | Number of Doctors | 1091 | | | 1171 | | | 1265 | | | |
| 986 | Bandung, Kota | Number of hospitals | 25 | | | | | | 29 | | | |
| 987 | Bandung, Kota | Number of Midwives | 357 | | | 417 | | | 537 | | | |
| 988 | Bandung, Kota | Number of people employed | | | 915047 | 952752 | 998227 | 948124 | 1012946 | 1067721 | 1055422 | |
| 989 | Bandung, Kota | Number of people employed in agriculture, forestry and fishery | | | 10652 | 13705 | 10996 | 9778 | 3381 | 10524 | 7460 | |
| 990 | Bandung, Kota | Number of people employed in construction sector | | | 55903 | 31476 | 42083 | 47858 | 39887 | 51259 | 58444 | |
| 991 | Bandung, Kota | Number of people employed in electricity and utilities sector | | | 702 | 4770 | | 3307 | | 6090 | 17840 | |
| 992 | Bandung, Kota | Number of people employed in financial services sector | | | 36506 | 45545 | 38750 | 38959 | 55035 | 72117 | 56381 | |
| 993 | Bandung, Kota | Number of people employed in industrial sector | | | 180680 | 194353 | 198714 | 174509 | 251166 | 263893 | 259268 | |
| 994 | Bandung, Kota | Number of people employed in mining and quarrying sector | | | | 1516 | 1860 | 1947 | 3762 | 3536 | 6401 | |
| 995 | Bandung, Kota | Number of people employed in social services sector | | | 181167 | 220646 | 231828 | 261553 | 230375 | 210292 | 199134 | |
| 996 | Bandung, Kota | Number of people employed in trade, hotel and restaurant sector | | | 367605 | 383787 | 370811 | 346110 | 369161 | 378926 | 392918 | |
| 997 | Bandung, Kota | Number of people employed in transportation and telecommunication sector | | | 81832 | 63899 | 103185 | 64103 | 60179 | 71084 | 57576 | |
| 998 | Bandung, Kota | Number of people in labor force | | | 1095616 | 1124411 | 1151180 | 1079477 | 1129744 | 1177291 | 1185474 | |
| 999 | Bandung, Kota | Number of people live below the poverty line (in number of people) | 84600 | 95200 | 87200 | 106800 | 110300 | 118600 | 116900 | 111100 | 117700 | 114999 |
| 1000 | Bandung, Kota | Number of people underemployed | | | 93000 | 121812 | 85370 | 114175 | 95651 | 109307 | 109678 | |
| 1001 | Bandung, Kota | Number of people unemployed | | | 180569 | 171659 | 152953 | 131353 | 116798 | 109570 | 130052 | |
| 1002 | Bandung, Kota | Number of Polindes (Poliklinik Desa/Village Polyclinic) | 1 | | | | | | | | | |
| 1003 | Bandung, Kota | Number of Puskesmas and its line services | 80 | | | | | | 73 | | | |
| 1004 | Bandung, Kota | Number of schools at junior secondary level | 242 | | | 116 | | | 244 | | | |
| 1005 | Bandung, Kota | Number of schools at primary level | 804 | | | 502 | | | 837 | | | |
| 1006 | Bandung, Kota | Number of schools at Senior Secondary level | 202 | | | 8 | | | 232 | | | |
| 1007 | Bandung, Kota | Number of Student: Junior Secondary Level (in number of people, 2009 data only) | | | | | 17031 | | | | | |
| 1008 | Bandung, Kota | Number of Student: Primary Level (in number of people, 2009 data only) | | | | | 210793 | | | | | |

Figure. 1. Open Data datasets preview

Table. 1. The indicator used for classification and clustering model constructions

| Variable | Years | | |
|---|---|---|---|
| | 2010 | 2011 | 2012 |
| *Human Development* | √ | | √ |
| *Gross Domestic Product* | √ | | √ |
| *Number of Population in* | √ | | |
| *Number of Internet* | √ | | |
| *Number of Labors* | √ | | |
| *Number of Population* | √ | | |

For classification model, we use Human Development Index (HDI), Gross Domestic Product (GDP), Number of Population in Poverty (NPP), Number of Internet Users (NIU), Number of Labors (NL), Number of Population (NP). These indicators have complete data only in 2010, the reason is because the census to collect some indicator data conducted every 10 years. Meanwhile for the clustering model, we use HDI and GDP. The objective is finding new characteristics whether there are places with high HDI even though they have low GDP or the other way around. The clustering model is useful to uncover hidden pattern behind the data.

We use 2012 data for clustering model, because the factor completeness and novelty of published data comparing to 2010 data. For the information, indicator NPP, NIU, NL, NP are constructor of HDI indicator. NIU or Number of Internet Users reflect whether a country considered as a developing or a developed country (Pratama and Al-Shaik, 2012). The proper use of Internet usage has positive correlation with HDI (Ssewanyana, 2011). NP and NL affected the NPP (Tambunan, 2011) (Hardini, 2011) (Mirza, 2012).

## 3. Model Construction and Result

In this section, we show the workflow to construct Classification model based on ANN and Clustering model based on K-means in Figure 3. Comparing to other famous classification based model such as Decision Tree, K-nearest neighborhood, Naïve Bayes, and others, ANN is chosen because the nature of the data and dynamic indicators proportion to predict HDI value.

After we collect the data, both of models need to do preprocessing data. For Classification, we get the best ANN model by iterate for 10 times for each the numbers of neurons in hidden layer with variation of 10 neurons, 13 neurons, 16 neurons, and 20 neurons to get the best model with the lowest mean error, which will be the best classification model. We perform classification to predict which city that fall into 4 categories of HDI. According to UNDP (Human Development Report, 2010), they are Low HDI, Medium HDI, High HDI, Very High HDI. We predict HDI class using 5 other data as the predictors/inputs using the best ANN Classification model. The final stage, we perform prediction to measure the model performance. The ANN visualization can be seen in Figure 4.

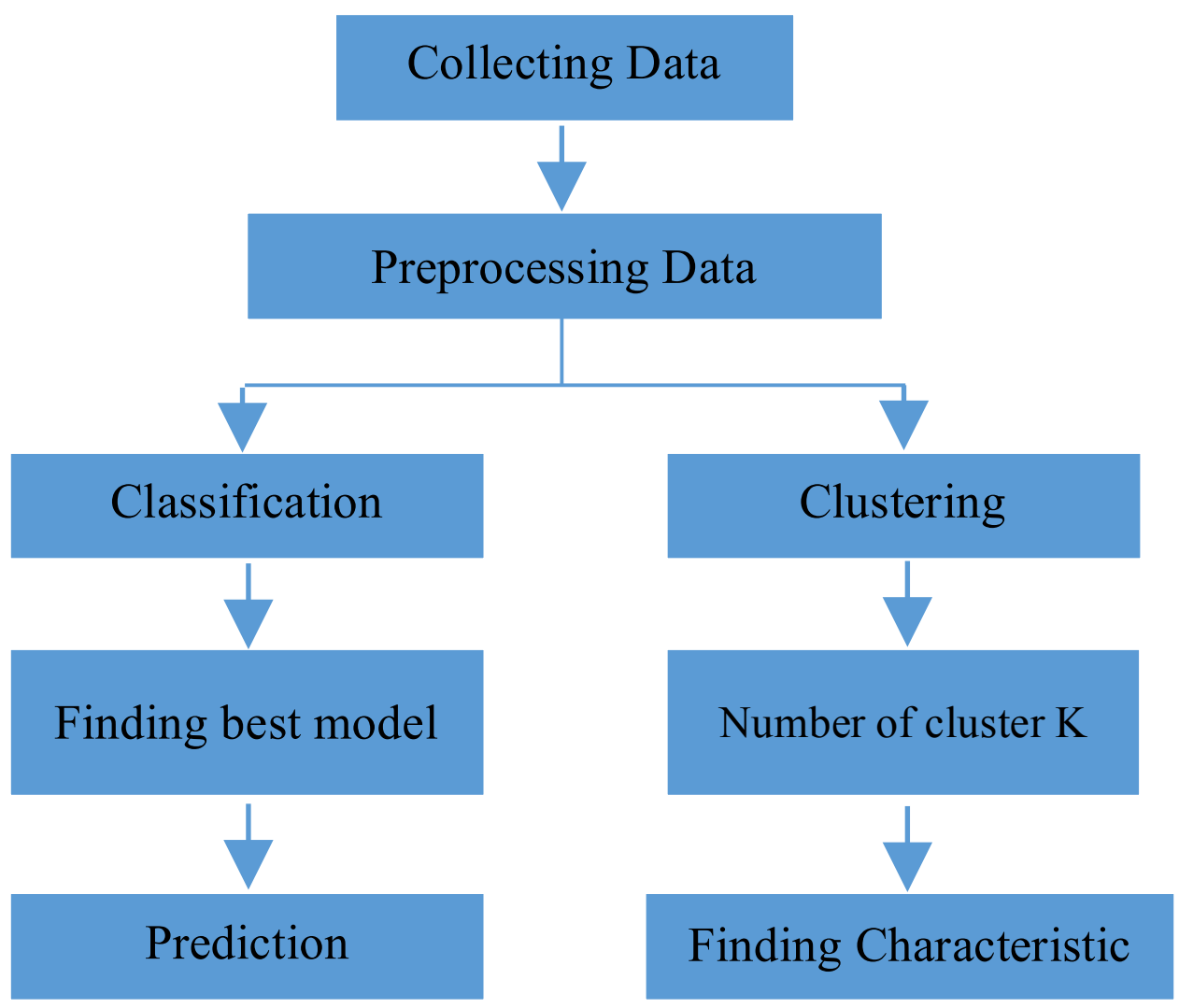

Figure 3. Research Process Flowchart

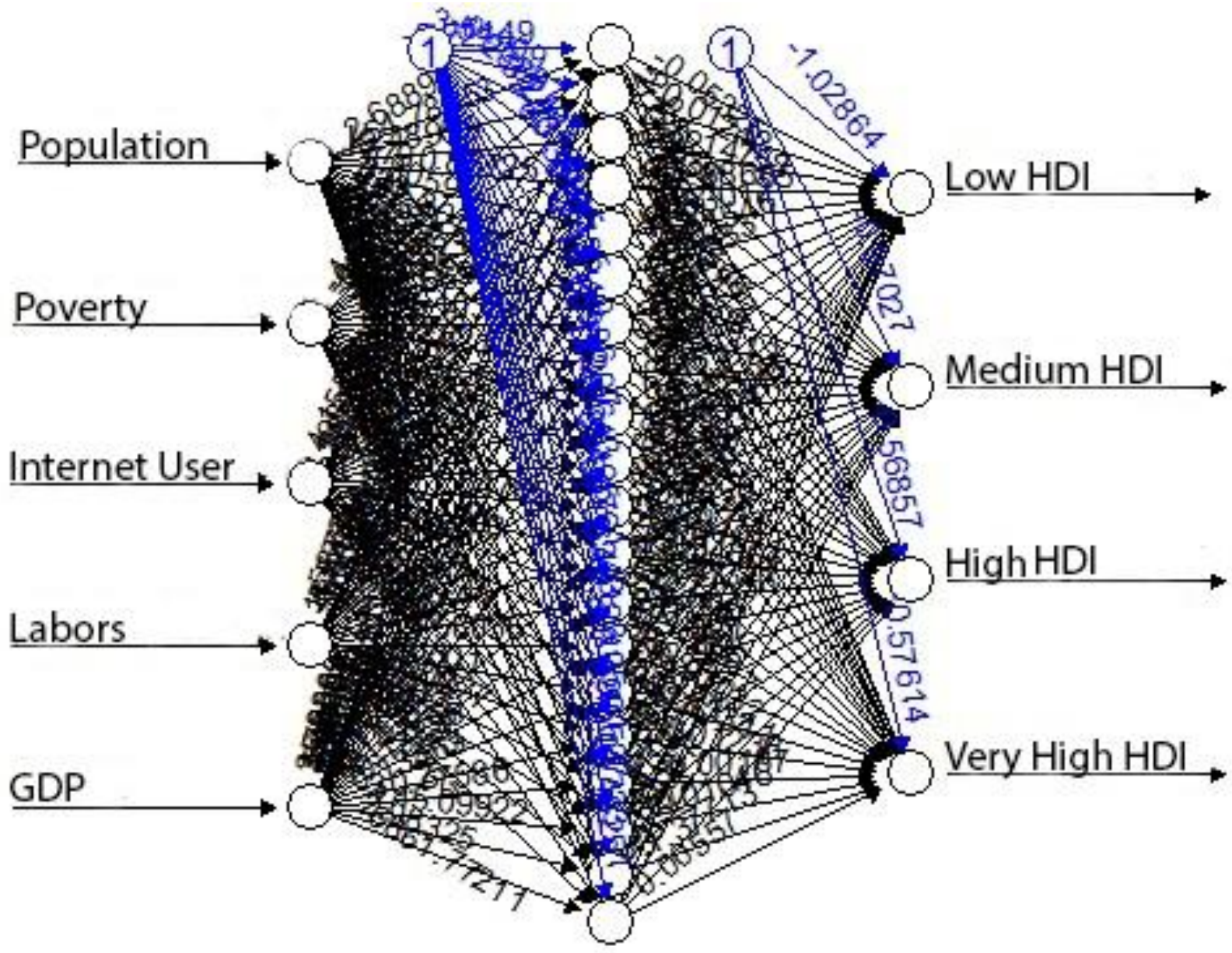

Figure 4. The result of Artificial Neural Network

Our ANN classification model construction result is the lowest mean error is 7.6596, which coming from the highest number of neurons in the experiment, which is 20 neurons. The mean error for 10 neurons, 13 neurons, and 16 neurons models are 10.53145, 12.12073, 8.380923 respectively.

Clustering model use HDI and GDP to find interesting clustering pattern between both indicators. We do lot of preprocessing step in this step, which are data noise removal, and transform into appropriate data format. The final data consists of 495 cities in Indonesia. K-means clustering calculations HDI and GDP begins by determining the number of $k$ cluster. In this study, we use $k = 4$ based on HDO and GDI Indonesia cities in 2012.

Looking at Figure 5. Cluster 3 has average HDI value of 72.39, which belong into the High HDI category. Cluster 1 has average HDI value of 52.30, which belong to Low HDI category. Cluster 4 has average HDI value of 76.82, which belong to High HDI category. At last, cluster 2 has average HDI value of 67.80, which belong to Medium HDI category. The HDI range value is distinctively separated between clusters, while in some GDP range value is overlap between clusters, especially when GDP value is below 40.

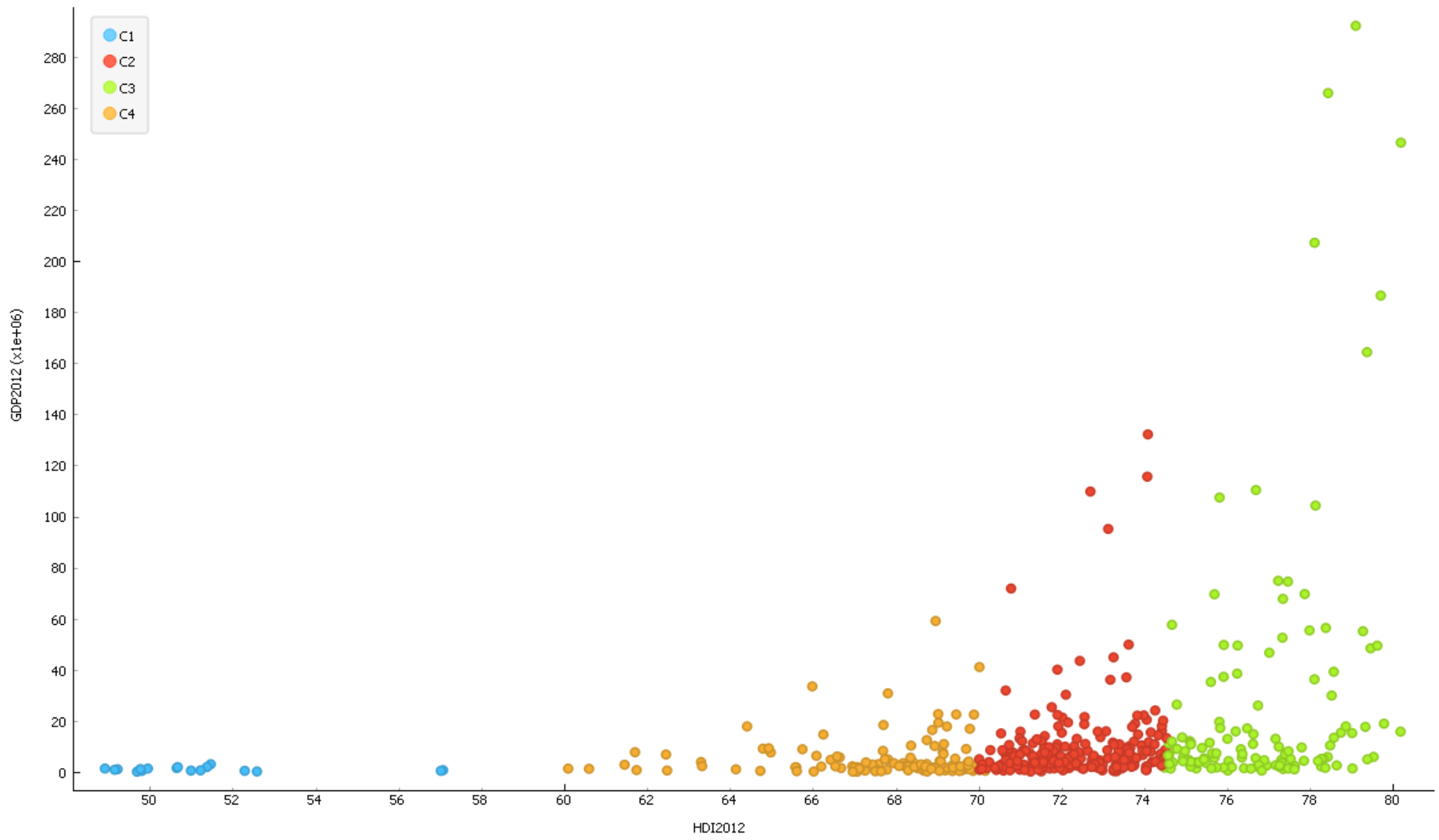

Figure 5. The Result of K-Means clustering model of HDI and GDP data in 2012. C1 is cluster 1, C2 is cluster 2, C3 is cluster 3, and C4 is cluster 4.

## 4. Model Evaluation, Analysis, and Conclusion

To evaluate the accuracy of classification model we use confusion matrix, which basically a matrix to describe the performance of classifier. The confusion matrix can be seen in Figure 5. The result is from 99 data, we predicted 90 data is correctly predicted and 9 misclassification data. The ANN classification model with 5 inputs, 20 neurons, and 4 outputs configuration or (5:20:4) have 9.09% prediction error. For the K-means clustering model evaluation based on 4 clusters construction, we are able to predict all new data into the right cluster. In short, we have 100% accuracy for clustering model.

| | **High Human Development** | **Medium Human Development** | **Low Human Development** |
|---|---|---|---|
| **High Human Development** | **88** | **1** | **0** |
| **Medium Human Development** | **7** | **2** | **0** |
| **Low Human Development** | **1** | **0** | **1** |

Figure 5. The confusion matrix to measure the performance of ANN classification model

For the analysis, Classification models able to classify the HDI status of any Indonesia city with high accuracy based on 5 indicators: GDP, NPP, NIU, NL, and NP. In practical usage, we can learn HDI class from the value of 5 indicators. In some cases, we can predict future HDI value in real time based on today indicators value. Clustering models able to separate different cluster characteristics based on HDI and GDP indicators. In Figure 5. show the different characteristic of each cluster. By this visualization, we know exact the condition of each city by their HDI and GDP value. For example, cities in cluster 3 and 4 have High HDI value that means they good value of health aspect and life expectancy, education aspect, and income aspect. In those clusters, most cities have low level of GDP, only handful of city have high level of GDP. This shows that HDI not only determined by GDP value, but mostly because of the regional government efficiency to make high HDI value happen.

The conclusion is that by having a good and systematical effort to support *Open Data* movement to collect rigorous data, then citizen and government can evaluate the government program or policy to boost any government project to increase citizen welfare. From the HDI case study, we learn that we are able to make such predictions even with the condition of limited data. The possibility of having many models is unlimited with the availability of complete data supported by *Open Data* movement. We can perform deeper, complex analysis, and verification-examination by different model available. In the end, government will have unified voice based on data analytical process in making policy or program.